\documentclass[prl,twocolumn,amsmath]{revtex4}
\usepackage{epsfig}
\usepackage{amssymb}
\usepackage{amsmath}
\usepackage{graphicx}

\begin{document}

\title{Quantum Conditions on Dynamics and Control in Open Systems}
\author{Lian-Ao Wu, Arjun Bharioke and Paul Brumer}
\affiliation{Center for Quantum Information and Quantum Control, and Chemical Physics
Theory Group, Department of Chemistry, University of Toronto, 80 St. George
Street, Toronto, Ontario M5S 3H6, Canada}

\begin{abstract}
Quantum conditions on the control of dynamics of a system coupled
to an environment are obtained. Specifically, consider a system
initially in a system subspace $H_{0}$ of dimensionality $M_{0}$,
which evolves to populate system subspaces $H_{1}$, $H_{2}$ of
dimensionality $M_{1}$, $M_{2}$. Then
there always exists an initial state in $H_0$ that does not evolve into $H_2$ if $%
M_{0}>dM_{2},$ where $2 \leq d \leq (M_0 +M_1 +M_2)^2$ is the
number of operators in the Kraus representation. Note,
significantly, that the maximum $d$ can be far smaller than the
dimension of the bath. If this condition is not satisfied then
dynamics from $H_{0}$ that avoids $H_{2}$ can only be attained
physically under  stringent conditions. An example from molecular
dynamics and spectroscopy, i.e. donor to acceptor energy transfer,
is provided.
\end{abstract}

\maketitle

The study of open quantum systems\cite{alicki,petrucionne,lendi} is of great current interest, a
consequence of the
desire to understand and manipulate devices that are reliant
on quantum effects. Of particular interest is the control of such
quantum systems, with coherent control\cite{Brumer03,Rice00} being a
most promising approach.

The essential principle of the coherent control of atomic and molecular
processes relies on the creation of multiple interfering pathways to the same final
state \cite{Brumer03,Rice00}. Manipulation of laboratory parameters that
affect these coherent pathways then allows direct control over the
associated interference contributions,
and hence control over the relative cross sections
involved in the process. Numerous theoretical scenarios and experimental
results, in a host of applications, have successfully demonstrated such
control.

Despite the importance and generality of the field, there are remarkably few
formal results that provide necessary and sufficient conditions under which
complete control of the system dynamics is possible. These include theorems
such as that of Huang-Tarn-Clark\cite{HTC1983}, a theorem by Ramakrisnan $et$
$al$. on the dimensionality of the Lie Algebra induced by the interaction
between the system and the control field \cite{Rama1995}, and a theorem by
Shapiro and Brumer\cite{Brumer95}, where control was shown to depend on the
dimensionality of the controlled subspaces. These theorems all deal with
closed systems, i.e. those that are not in contact with an environment. Real
quantum systems are, however, not isolated. They often interact with a
surrounding environment, which is regarded as an uncontrolled system of
arbitrarily large dimension. As a result the system is open to its
environment, and proofs regarding control in  open systems are becoming the focus of
serious attention. For example, Reference \cite{Rabitz07} provides formal
theorems on controllability in open system dynamics treated via the Kraus
representation.

In this letter we significantly generalize a prior
argument\cite{Brumer95} for closed systems and obtain quantum
control conditions in open systems. We show that the condition for
a specific control problem in open systems is stronger than that
in closed systems, but that the relative dimensionality of the
different subspaces remains a crucial determinant of
controllability. In addition, our approach gives the solution to
the control problem when it is achievable, and provides a
constructive method for extending various closed system proofs to
the open system case.

To expose the difference between the open and closed cases, to adopt a
common notation, and to provide a generalized formalism, we first reconsider
the closed system\cite{Brumer95}. Consider a system whose Hamiltonian
eigenstates are partitioned into bases for three subspaces: $H_{0}$, $H_{1}$%
, $H_{2}$, of dimensionality $M_{0}$, $M_{1}$ and $M_{2}$, respectively. Let
the basis vectors $\left\vert i,n_{i}\right\rangle $ span the subspace $%
H_{i},$ where $i=0,1,2$ labels the subspace and $n_{i}=1,...,M_{i}$ labels
the states in $H_i$. The total dimension of the system is $%
M=M_{0}+M_{1}+M_{2}.$ A superposition of basis vectors from
$H_{0}$ is chosen as the initial state. We assume that the system
initially resides in $M_0$ and that the state can flow into the
other two subspaces under the dynamics. The question under
consideration is: under what conditions can one prevent dynamics
into the subspace $H_{2}$, by preparing states in $H_0$ that go
solely into $H_1$? To resolve this problem, Ref. \cite{Brumer95}
considered linear superpositions of states in $H_0$ and utilized
interference between the resultant paths to $H_1$ and $H_2$ in
order to control the dynamics.

Using the evolution operator, $U$ we have:
\begin{equation}
\left\vert \psi \right\rangle =U\left\vert \psi _{0}\right\rangle
\label{eq1}
\end{equation}%
where $|\psi _{0}\rangle $ is the initial superposition in $M_{0}$ and $%
|\psi \rangle $ denotes the final state. With $U$ represented as an $M\times
M$ dimensional matrix, and $\left\vert \psi \right\rangle $ expressed as an $%
M$ column matrix we can examine the number of simultaneous equations that
need to be solved for $\left\vert \psi \right\rangle $ to have zero
population in the rows corresponding to $H_{2}$. Using this approach, Ref.
\cite{Brumer95} obtained conditions on the relative dimensionality of the
different spaces by requiring%
\begin{equation}
\left\langle 2,k_{2}\right\vert U\left\vert \psi _{0}\right\rangle =0
\label{eq10}
\end{equation}%
for all states $k_2$ in $H_2$.
Specifically, they showed that if $M_{0}>M_{2}$, it is always possible to
prevent transitions from $H_{0}$ into $H_{2}$ . By contrast, a large number
of difficult-to-satisfy linear dependence conditions are required to prevent
transitions if $M_{0}\leqslant M_{2}$. Significantly, the result only
depends on the dimensionality $M_{0}$ and $M_{2}$ of these two subspaces.

Consider then the same system, but where it is now open to an
environment, which can be either finite or infinite dimensional,
and that may, or may not, be in the presence of external fields.
As is typically the case, we assume that the system can be
addressed (e.g. via laser excitation) to prepare the initial
state, but that the environment remains unaffected. Further, we
assume that the initial system-bath density matrix $\rho
_{tot}(0)=\rho (0)\otimes \rho _{B}(0)$ is a separable product of
the initial system $\rho (0)$, usually a pure state that can be
selectively prepared, and initial bath $\rho _{B}(0)$ density
matrix\cite{foot}. The open system after evolution can be
characterized by an $M\times M$ reduced density matrix $\rho $,
defined as $\rho =$Tr$_{B}\rho _{tot}$, where $\rho _{tot}$ is the
final time total system-bath density matrix and Tr$_{B}$ indicates
a trace over the bath. Focusing on the reduced density matrix
(i.e. $\rho _{tot}$ traced over the bath) allows us to continue to
use the concept of dimensionality in an open system. In general,
the reduced density matrix evolves according to a linear
transformation
\begin{equation}
\rho =\sum_{\alpha ,\beta }A^{\alpha }\rho (0)B^{\beta }  \label{eq11}
\end{equation}%
For instance, in the case of natural quantum evolution, denoting the
propagator for the full (system + bath) as $U$, we have that
\begin{equation}
\rho _{tot}=U\rho _{tot}(0)U^{\dagger }=U\rho (0)\otimes \rho
_{B}(0)U^{\dagger },  \label{evolverho}
\end{equation}%
where we have used the separability of $\rho _{tot}$ at time zero. In
general $\rho _{B}(0)=\sum_{a,b}p_{ab}|a\rangle \langle b|$, where $%
|a\rangle $ and $|b\rangle $, and the $|e\rangle $ below, are eigenstates of
the bath Hamiltonian. By tracing Eq. (\ref{evolverho}) over the bath
coordinates, we have that
\begin{equation}
\rho =\sum_{e,a,b}A^{a,e}\rho (0)B^{a,b,e}  \label{eq20}
\end{equation}%
with $A^{a,e}=\langle e|U|a\rangle $ and $B^{a,b,e}=p_{ab}\langle
b|U^{\dagger }|e\rangle $.

Under the assumption of complete positivity \cite%
{Nielsen00}, Eq. (\ref{eq11}) simplifies and the reduced density
matrix evolves according to the operator-sum representation (i.e.
the canonical Kraus representation), which is generic physically for the
initial product density matrix \cite{lendi}:
\begin{equation}
\rho =\sum_{\alpha =1}^{d}E^{\alpha }\rho (0)E^{\alpha \dagger }  \label{eq2}
\end{equation}%
Here the sum over Kraus operators\cite{Nielsen00} $E^{\alpha}$ is
such that $\sum_{\alpha =1}^{d}E^{\alpha \dagger }E^{\alpha }\leq
I$. Specifically, $\sum_{\alpha =1}^{d}E^{\alpha \dagger
}E^{\alpha }=I$ for trace-preserving quantum operations and
$\sum_{\alpha =1}^{d}E^{\alpha }E^{\alpha \dagger } < I$ for
non-trace-preserving quantum operations such as quantum
measurements. Note, significantly, that the operator-sum
representation has at most $M^{2}$ Kraus operators, i.e., $2<d
\leq M^{2}$ and that the operators are fixed by the given system, bath
and any incident external fields.

We continue below to consider
the general form [Eq. (\ref{eq11})], but subsequently focus on
evolution under the Kraus representation.

Note first that Eq. (\ref{eq11}) means that the operator $\rho (0)$ can be
linearly transformed into the operator $\rho .$ If we denote the operation
connecting the matrix $\rho (0)$ to $\rho $ by $V,$ then $V$, in a
particular representation, is a matrix with four subscripts. That is,
\begin{equation*}
\rho _{ts}=\sum_{t^{\prime }s^{\prime }}V_{tst^{\prime }s^{\prime }}\rho
_{t^{\prime }s^{\prime }}(0)
\end{equation*}%
where $t$ or $s$ denotes the indices $(i,n_{i})$ of the basis vectors $%
\left\vert i,n_{i}\right\rangle .$ Rewriting $\rho $ as $M^{2}$ dimensional
column vector denoted $\widetilde{\rho }$, Eq. (\ref{eq2}) can be
equivalently rewritten as:
\begin{equation}
\widetilde{\rho }=\widetilde{V}\widetilde{\rho }(0)  \label{eq3}
\end{equation}%
where the explicit form of the \ $M^{2}\times M^{2}$ matrix $\widetilde{V}$
is
\begin{equation*}
\widetilde{V}=\sum_{\alpha ,\beta }A^{\alpha }\otimes B^{\beta }
\end{equation*}

Equation (\ref{eq3}) is seen to have the same form as Eq. (\ref{eq1}),
allowing the approach used earlier for a closed system to be extended to
open systems. Consider then an initial reduced density matrix that is in $%
H_{0}$, i.e.
\begin{equation*}
\rho (0)=\left(
\begin{array}{ccc}
\rho _{M_{0}^{2}} & 0 & 0 \\
0 & 0 & 0 \\
0 & 0 & 0%
\end{array}%
\right) ,
\end{equation*}%
where $\rho _{M_{0}^{2}}$ is an \ $M_{0}\times M_{0}$ matrix. In order to
prevent dynamics from going into $H_{2},$ we require
\begin{equation}
\rho _{(2,k_{2})(2,k_{2})}=0  \label{eq30}
\end{equation}%

For the case of interest the Kraus representation applies so that a
propagated initial pure state
$\rho _{M_{0}^{2}}=\left\vert
\psi _{0}\right\rangle \left\langle \psi _{0}\right\vert $ ,
becomes
\begin{equation*}
\rho _{(2,k_{2})(j,n_{j})}=\sum_{\alpha =1}^{d}\left\langle
2,k_{2}\right\vert E^{\alpha }\left\vert \psi _{0}\right\rangle \left\langle
\psi _{0}\right\vert E^{\dagger \alpha }\left\vert j,n_{j}\right\rangle
\end{equation*}%
Given that the
diagonal element $\rho _{(2,k_{2})(2,k_{2})}$ is $\sum_{\alpha
=1}^{d}\left\vert \left\langle 2,k_{2}\right\vert E^{\alpha }\left\vert \psi
_{0}\right\rangle \right\vert ^{2}$, satisfying $\rho _{(2,k_{2})(2,k_{2})}=0$ means
requiring%
\begin{equation}
\left\langle 2,k_{2}\right\vert E^{\alpha }\left\vert \psi _{0}\right\rangle
=0  \label{eq40}
\end{equation}%
This being the case, demanding zero population in $H_2$ also
implies that all elements
$\rho _{(2,k_{2})(j,n_{j})}$, which include coherences with the $H_2$ Hilbert
subspace, are also zero.

Equation (\ref{eq40}) has a nontrivial solution if
$M_{0}>dM_{2}$, and control via initial state
preparation is achievable no matter what the form or dynamics of
the $d$-dimensional Kraus operators. Further, this equation provides the initial
state that allows for the desired control. However, this control
condition is, as expected, far more stringent in the open system
than in the closed system case, where $d=1$. Note, however, that
the bath effects are still limited, The minimum $d$ can
be\cite{alickiprivate}  as small as two, and the maximum $d$ is
$M^2$, which can be far smaller than the dimensionality of the
bath.

If $M_{0} \le dM_2$, control via initial state preparation is also
far more difficult in the open system than in the (already
difficult) closed system case. Define a $dM_{2}\times M_{0}$
dimensional matrix $W$ with matrix elements $W_{(2k,\alpha
),(0n)}=$ $\left\langle 2,k\right\vert E^{\alpha }\left\vert
0,n\right\rangle .$ In the case where $M_{0}\leqslant dM_{2}$, the
rank of $W$ is equal to $M_{0}$ unless all $M_{0}\times M_{0}$
dimensional submatrices of $W$ are singular. Hence, non-trivial
solutions to Eq. (\ref{eq40}) exist if
det$(W_{M_{0},M_{0}}^{(k)})=0$, where $(k)$ numbers all of the
submatrices. This condition also implies that a set of columns of
$W$ are linearly dependent. Hence control is possible for a 
{\it specific} class of  $W$, expected to be difficult to obtain physically,

Note that the definition of the subspaces $H_i$ and associated dimensions
$M_i$ can, in some instances, also be manipulated. For example, in
bound state systems (such as that in the example below) these subspaces
can be defined by accessing only specific system eigenstates using a pulsed
laser field. This facility might prove additionally useful in attempting to
satisfy the $M_0 > d M_2$ bound of this theorem.

These results allow numerous applications. Consider, for example, electronic
energy transfer from donor to acceptor molecules, where both may be part of
one larger molecule. These systems are
ubiquitous, ranging from relatively small systems\cite{Levy} to large
structures such as carotenoid-to-bacteriochlorophyll
energy transfer in photosynthesis\cite{Scholes}.
The most interesting cases take place in condensed matter environments, i.e.
open systems. Studies of the dynamics
and spectroscopy of such systems can be carried out by laser excitation of
the system from a lower electronic manifold (here $H_0$) to the donor (here $H_1$).
Electronic energy transfer from the donor ($H_1$)
to the acceptor ($H_2$) is then measured.
Here, the subspaces $H_i$, and associated dimensions $M_i$,
are determined by a combination of the molecular state densities and the
width of the laser pulses
that prepare a preliminary superposition of states on $H_0$
and that subsequently excite the system into $H_1$ and $H_2$.

In some systems of interest, excitation from $H_0$ to the donor is
contaminated by partial excitation of the acceptor as well, with a
concomitant reduction in the quality of the data on the subsequent
electronic energy transfer dynamics. A considerable improvement
would result from being able to excite $H_1$ with reduced
population transfer to $H_2$ from $H_0$. Results of the open
system theorem indicate that this is (a) difficult to achieve
physically if $M_0 \le d M_2$, and (b) attainable if $M_0 > d
M_2$.  In the latter case, one could ensure significantly reduced,
and ideally zero, acceptor population at the target time. The
extent to which this is achievable is dependent upon the
particular system; specific systems of this type will be the
subject of future study.

In summary, the quantum conditions obtained above are completely
general, applying to both systems that are controlled, as well as
to uncontrolled system evolution. The result establishes an
important inequality between the dimension of the subspace $M_2$
which we desire not to populate, the initial subspace $M_0$ from
which dynamics evolves, and the dimensionality of the Kraus
representation.  It is a dynamics-independent property of $d$-dimensional Kraus
evolution.  As long as the geometric condition $M_0  \geq dM_2$ is
satisfied, control via initial state preparation is achievable.
Further, the fact that control can be achieved via initial state
preparation, in the presence of an environment under the
prescribed conditions, is useful for control applications in
realistic systems.

Two  supplementary remarks are in order.
First, this theorem places emphasis on the
importance of the dimensionality $d$ of the Kraus representation,
known to satisfy $2 \leq d \leq (M_0 +M_1 +M_2)^2$. Hence, these
results should motivate further studies to determine the actual $d$ for
realistic systems.
Second,
 we note that an analogous method to that described
above can be used to extend other closed system proofs, such as those of
Refs. \cite{HTC1983} and \cite{Rama1995}, to open systems.
\newline

\textbf{Acknowledgments} We thank Professor R. Alicki
for discussions on Kraus operators, and
Professor G.D. Scholes for comments on donor-acceptor dynamics.
This work was supported by NSERC Canada.


\end{document}